# A Two Wire Waveguide and Interferometer for Cold Atoms

E. A. Hinds, C. J. Vale, and M. G. Boshier

*Sussex Centre for Optical and Atomic Physics,
University of Sussex, Brighton, BN1 9QH, U.K.*

A versatile miniature de Broglie waveguide is formed by two parallel current-carrying wires in the presence of a uniform bias field. We derive a variety of analytical expressions to describe the guide and present a quantum theory to show that it offers a remarkable range of possibilities for atom manipulation on the sub-micron scale. These include controlled and coherent splitting of the wavefunction as well as cooling, trapping and guiding. In particular we discuss a novel microscopic atom interferometer with the potential to be exceedingly sensitive.

An atom whose magnetic moment has projection $\mu_\zeta$ along an external magnetic field of magnitude $B$ experiences a Zeeman interaction potential $U = -\mu_\zeta B$. The associated force is able to guide weak-field-seeking atoms along a minimum of $B$. This principle underlies the Stern-Gerlach effect and the magnetic hexapole lens, which have played such important roles in the history of atomic beams. Magnetic forces are now a central feature of atom optics – the subject of manipulating, confining, and guiding cold neutral atom clouds and Bose-condensates [1, 2, 3].

Recently there has been great interest in building miniature magnetic structures where small features make a strong field gradient, while a superimposed uniform bias field makes a zero whose position is adjustable [4]. This idea has been realized in several laboratories, using either supported wires [5, 6, 7], printed circuits [8, 9, 10, 11, 12] or microscopic patterns of permanent magnetization [13]. Miniature guides are attractive because they offer the possibility of propagating de Broglie waves in a single transverse mode in 1D [14] or 2D [15]. This is a central goal of many experimental groups because it is required for achieving atom interferometry with guided de Broglie waves. Miniature structures are also promising for studying the physics of quantum gases confined to less than 3D [16]. In this letter we discuss the guide formed by two wires carrying parallel currents $I$ spaced $2A$ apart in the presence of a bias field, as illustrated in Fig. 1. We point out that the guide is far more adaptable than previously realized [9,10], deriving simple formulae for the various configurations it can produce. We present a quantum theory to show how the guide can be used to manipulate atoms on the sub-micron scale, to make a controlled and coherent splitting of the wavefunction, and to realize a novel atom interferometer.

Let us adopt the natural units of $A$ for lengths, $B_0 \equiv \mu_0 I / 2\pi A$ for magnetic fields, and $\mu_\zeta B_0$ for energies. Dimensionless quantities based on these units are indicated by lower case letters, e.g. $x = 1$ means $X = A$. With this scaling the magnetic field produced by the current-carrying wires in Fig. 1 has Cartesian components

$$b_x = \frac{-y}{(1+x)^2 + y^2} + \frac{-y}{(1-x)^2 + y^2}$$
$$b_y = \frac{1+x}{(1+x)^2 + y^2} + \frac{-1+x}{(1-x)^2 + y^2}$$
(1)

On the y-axis, $b_y = 0$ and the field produced by the wires is $b_x$. This has a single maximum of $b_x = 1$ at height $y = 1$ as shown in Fig. 2. The addition of a bias field $\beta < 1$ (in normalized units) along the positive x-direction cancels $b_x$ at two positions, indicated for $\beta = 1/2$ by the circles in Fig. 2. For weak-field-seeking atoms, these zeros form guides parallel to the z-axis (a small field may be added along the z-direction to suppress non-adiabatic spin flips, although states without angular momentum around the guide axis can be stable without it [14]). The inset in Fig. 2 shows the dimensionless guiding potential $u$. The barrier between the two guides is $1 - \beta$, while the barrier above the upper guide is $\beta$. As the strength of the bias field is increased, these zeros approach one another until they coalesce to form a single guide at height $y = 1$ when $\beta = 1$. Further increase of the bias splits the guide horizontally and the zeros follow the circle $x^2 + y^2 = 1$ (Fig. 1). In the first two rows of Table 1 we present simple formulae giving the guide centers $(x_0, y_0)$ and trap depths $u_0$ in each of the three bias field regimes.

Fig. 3a shows the interaction potential $u$ and the field lines for $\beta = 0.8$. The trap centered on $y = 1/2$ is smaller and 4 times stronger than the upper one at $y = 2$ (the ratio of gradients is always the inverse of the ratio of heights). Each guide is cylindrically symmetric over a limited region around its center, and has a constant potential gradient, a characteristic of the quadrupole symmetry evident in the field lines. We note that the fields are oppositely directed in the two guides. At the two bottom corners, the field can be seen circulating around the current-carrying wires, which are taken in this diagram to be of negligible radius. Figure 3b shows the potential and field lines for the single guide, formed at the critical bias. Here the linear gradient vanishes: the guide is harmonic with curvature $\partial^2 u/\partial \rho^2 = 1$ and the corresponding hexapole symmetry can be seen in the field lines. At higher bias field the potential splits laterally into two quadrupole guides as illustrated in Fig. 3c for $\beta = 1.5$. Simple formulae for the gradients are given in the last row of Table 1.

If we allow the bias field to have a component in the $y$-direction, many other possibilities open up, one of which is particularly relevant, as we will see below. Let us apply both the critical bias $\beta = 1$ along $x$, and an extra bias of magnitude $\Delta \beta$ at angle $\vartheta$ to the $x$-axis. When $\Delta \beta \ll 1$, the potential splits into two guides separated by a distance $2\sqrt{2\Delta \beta}$, and the line joining their centers makes an angle $\vartheta/2$ with the $x$-axis. Thus, when the extra bias field is rotated the two guides orbit around the coalescence point at half the frequency.

Each of the three regimes of bias field has interesting features to offer for atom optics experiments. To take a concrete example, consider a guide with a 300 μm spacing and 2 A flowing in the wires, for which the characteristic field and gradient are $B_0$ = 2.7 mT and $B_0/A$ = 18 T/m (such a structure already exists in our laboratory). With a weak bias field of 0.3 mT ($\beta = 0.1$), the upper guide is centered 3 mm ($2A/\beta$) above the wires and has a field gradient of 0.1 T/m ($\frac{1}{2}\beta^2 B_0/A$). This is very well suited to operate as a magneto-optical trap (MOT) with two pairs of suitably polarized light beams in the *x-y* plane, propagating along axes rotated by 45º from the *x*- and *y*-axes.  For the purpose of initially collecting atoms in the MOT, a pair of auxiliary coils can create a field gradient along the *z*-direction, allowing a third pair of light beams to produce MOT confinement along the *z*-axis.  The MOT can be lowered and compressed to a maximum gradient of 4.4 T/m ($\frac{1}{4}B_0/A$) by increasing the bias field to 0.23 mT ($\sqrt{3}B_0/2$).  At this point the light can be turned off and the auxiliary coils producing the field gradient along the *z*-axis can be switched to produce parallel fields.  This makes a purely magnetic Ioffe-Pritchard trap, where weak-field-seeking atoms can be left in the ground state by evaporation. Alternatively, if a cold dense sample of atoms is already available in a macroscopic magnetic trap, as for example with a Bose-Einstein condensate, a single mode of the upper guide can be loaded in a "mode-matched" way by suddenly switching off the trap and turning on the guide with *I*, $\beta$ and a *z*-bias field chosen to duplicate the original trap spring constants.  In either case, a subsequent adiabatic variation of the field can bring atoms to the coalescence point of the guide, where the potential is harmonic and the transverse frequency $\omega_0$ is given by $m\omega_0^2 A^2 = \mu_\zeta B_0$.  This has the value $2.8 \times 10^3$ s$^{-1}$ for the particular guide we are considering here.  The ground state size $\sigma \equiv \sqrt{\hbar/m\omega_0}$ is 512 nm.

Suppose now that the atoms have all been prepared in the transverse ground state of the harmonic guide.  An increase $\Delta\beta$ of the bias field deforms the guide potential into a symmetric double well, providing a highly controlled and reproducible 50:50 coherent splitter for the de Broglie wave.  If the two halves of the atom cloud are held apart for some length of time they may act as the arms of an interferometer.  In order to understand the operation of this interferometer in more detail, we have calculated the eigenmodes of the guide numerically by solving the two-dimensional time-dependent Schrödinger equation with the spin degree of freedom adiabatically eliminated. We start with the eigenstates $\{(n_x, n_y)\}$ of a harmonic potential, then slowly deform the potential to obtain the corresponding eigenstates of the Hamiltonian $\mathcal{H}$ for the 2-wire guide at the critical bias $\beta = 1$. Next we slowly vary the bias field to find how the three lowest eigenstates and their energies $\langle \mathcal{H} \rangle$ evolve as a function of $\beta$. We assume that the atom density is low, although of course the mean field interactions at higher density should produce interesting non-linearities and physics beyond the Gross-Pitaevskii equation.

Figures 4 and 5 show the eigenstates and their energies, labeled by the quantum numbers $(n_x, n_y)$ which count the number of nodes along *x* and *y*. At the critical bias, the spectrum is (almost) harmonic with the (0,0) state lying $\hbar\omega_0$ below the nearly degenerate pair of first excited states (1,0) and (0,1) (a small anisotropy of the guide prevents exact degeneracy). An increase $\Delta\beta$ of the bias splits the guide and the single-peaked (0,0) state deforms adiabatically into a double-peaked wavefunction with even reflection symmetry in the *yz*-plane. As long as it is not perturbed, this of course returns to the harmonic ground state when $\beta$ is restored to 1. If, however, a differential phase of $\pi$ is introduced between its two peaks while $\beta > 1$, the wavefunction becomes a *yz*-antisymmetric function, which we recognize in Fig. 4 as the (1,0) state. When $\beta \to 1$ this evolves adiabatically into the first *yz*-antisymmetric excited state of the harmonic guide. For arbitrary phase shifts the final state of the harmonic guide is a superposition of (0,0) and (1,0). Thus symmetry dictates that the two output ports of the interferometer are two different vibrational states of the harmonic guide. The third state $(0,1)$ in Fig. 4 is even under reflection in the $yz$-plane, like the ground state, but it has a vibrational excitation in the $y$-direction. This excitation is preserved as the bias increases and consequently the energy lies approximately $\hbar\omega_0$ above $(0,0)$.

It is also interesting to consider decreasing the bias to $\beta < 1$. The interferometer states $(0,0)$ and $(1,0)$ both emerge in the upper guide, with $(1,0)$ having one quantum of excitation along the $x$-axis. The $(0,1)$ state however, goes into the ground state of the lower guide (Fig. 4).

Now we turn to the practical aspects of the interferometer. One can estimate the minimum $\Delta\beta$ needed to achieve splitting, $\Delta\beta_{min}$, by setting the displacement of the wells equal to the ground state diameter $2\sigma$. To lowest order the result can be expressed in the elegant form $\Delta\beta_{min} = 2\sigma^2 / A^2 = 2\hbar\omega_0 / \mu_\zeta B_0$, which for the guide in our example amounts to a change of 64 nT in the bias field. Achieving this level of control over the field would require some care but it is not a major technical challenge. When $\Delta\beta$ is increased further, the splitting between the $(0,0)$ and $(1,0)$ levels becomes exponentially small (Fig. 5), being equal to the tunneling frequency between the left and right potential wells. As the bias changes, one wants to avoid non-adiabatic transitions induced by $\partial \mathcal{H} / \partial t$. This operator, being symmetric under reflections in the $yz$-plane, cannot excite the $yz$-antisymmetric state $(1,0)$, but it does connect the ground state to $(0,1)$. Since this and the other coupled states are at least $\hbar\omega_0$ away in energy, the adiabatic condition is that $\Delta\beta$ must change slowly in comparison with the period of harmonic oscillation $1/\omega_0$ (over several milliseconds for the guide in our example). Once the atoms are split, the antisymmetric perturbation to be measured is turned on. As is usual in interferometry, this interaction has to be non-adiabatic to mix the states (1,0) and (0,0) (i.e. to introduce a phase shift $2\varphi$ between the left and right wavepackets) and must therefore be turned on and off in times much less than the tunneling period. This is not a stringent requirement since the tunneling can be made arbitrarily slow by a modest increase in $\Delta\beta$. Finally, reducing $\Delta\beta$ to zero, it remains only to read out the fringe pattern through the population in state (0,0) [(1,0)], which is proportional to $\cos^2\varphi$ [$\sin^2\varphi$].

In principle the readout can be done by absorption or fluorescence imaging to determine the atom distribution in the guide, although this method requires high spatial resolution. Alternatively, there are at least two methods for separating the (0,0) and (1,0) populations. First, if the guide is operated without any axial bias field and $\beta$ is reduced below 1 to extract both states in the upper quadrupole guide, the excited state (1,0) will be lost due to spin flips [14], leaving the (0,0) population to be measured by fluorescence. Second, the additional bias field $\Delta\beta$ can be rotated adiabatically from the $\hat{\mathbf{x}}$ direction to $-\hat{\mathbf{x}}$ before reducing its amplitude to zero, thereby transforming state (1,0) into (0,1). This leaves the output of the interferometer as a superposition of (0,0) and (0,1) in the harmonic guide, which can be read out by reducing $\beta$ so that the (0,0) component moves into the upper guide while the (0,1) part is transported downward.

In an interferometer the statistical signal:noise ratio is proportional to the interaction time $\tau$ and to the square root of the number of atoms $\sqrt{N}$. We can easily imagine $10^6$ atoms trapped in the waveguide with a measurement time of ~10 s, giving $\tau\sqrt{N} = 10^4$. In comparison a macroscopic cold atom interferometer of the Kasevich-Chu type [17] has $10^8$ atoms falling through the apparatus per second with an interaction time of order 30 ms, giving $\tau\sqrt{N} = 10^3$ over the same 10s. Both interferometers have a separation of ~100 μm between arms. It therefore seems clear that this method of splitting the atoms in time, rather than splitting atoms propagating through space, can significantly advance some kinds of measurement. For example, the new interferometer would be extremely sensitive to electric field gradients and to gravity. Also, it would not suffer from phase shifts due to unwanted rotations because the Sagnac phase is zero.

In conclusion, we have shown that two currents and a bias field form an exceedingly versatile structure, producing a waveguide that can be split in a highly controlled way and manipulated on the sub-micron scale. We have shown explicitly how a novel and sensitive atom interferometer could be realized with such a guide. This structure is also ideal for a variety of topical applications including a miniature magneto-optical trap and the study of 1-dimensional quantum gases. Finally, the quantum model of guided atom interferometry presented here will apply to many existing experiments once they reach the level of single-mode operation.

We are indebted to David Lau, Stephen Hopkins, and Mark Kasevich for valuable discussions and to Matthew Jones for experimental work on the Sussex microscopic guide. This work was supported by the UK EPSRC and the EU.

## Figure Captions

**Figure 1.** Atom guide using two current-carrying wires and a bias field. With increasing bias, two guiding regions move towards each other along $Y$ (dotted line) until they coalesce at $Y = A$. They then separate along the dashed circle.

**Figure 2**. Field $b_x$ on the x-axis versus height $y$ above the wires. A bias $\beta = 1/2$ along $x$ makes two zeros (circles). *Inset*: interaction potential $u$ showing the two linear guides.

**Figure 3**. Interaction potentials and field lines for (a) $\beta = 0.8$ (b) $\beta = 1$ (c) $\beta = 1.5$.

**Figure 4**. Wavefunctions of the lowest three states in the guide for extra bias fields $\Delta\beta / \Delta\beta_{min} = -2, 0, 2$. The ranges for $x$ and $y$ are $\pm 0.5$ and $0.5 - 1.72$ respectively.

**Figure 5**. Energy spectrum for the three lowest eigenstates in the guide versus extra bias field.

**Table 1**. Center, depth and gradient of the 2-wire guides for each regime of the bias field.

| Bias | $\beta < 1$ | $\beta = 1$ | $\beta > 1$ |
|---|---|---|---|
| $(x_0, y_0)$ | $\frac{1}{\beta}\left(0,\ 1\pm\sqrt{1-\beta^2}\right)$ | $(0,1)$ | $\frac{1}{\beta}\left(\pm\sqrt{\beta^2-1},\ 1\right)$ |
| $u_0$ | $\beta,\ 1-\beta$ | $1$ | $\beta-1$ |
| $\partial u/\partial \rho$ | $\sqrt{1-\beta^2}\left(1\pm\sqrt{1-\beta^2}\right)$ | $0$ | $\beta\sqrt{\beta^2-1}$ |

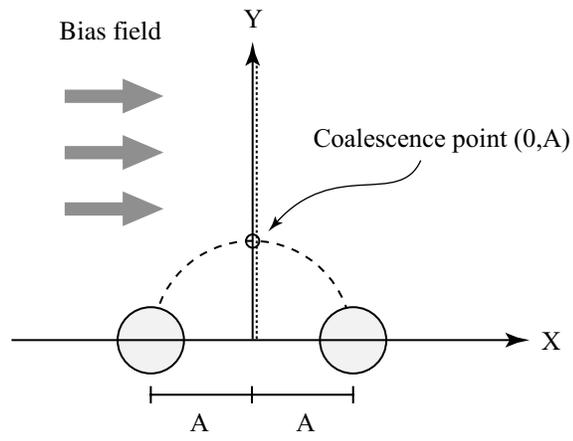

Figure 1: Hinds *et al*.

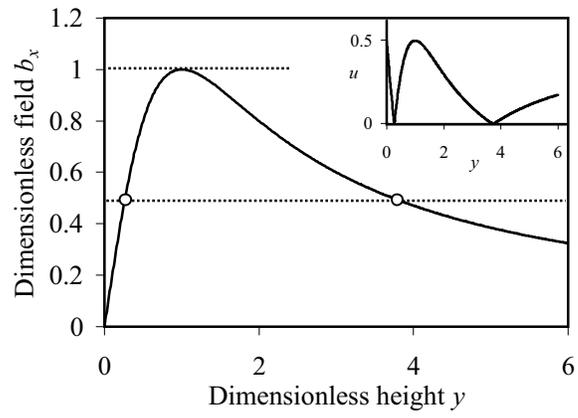

Figure 2: Hinds *et al*.

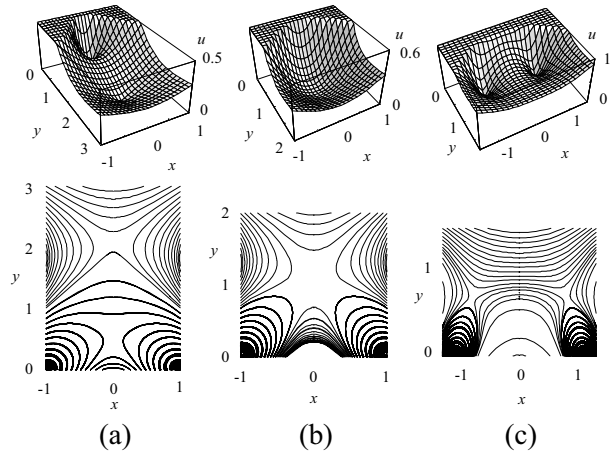

Figure 3: Hinds *et al.*

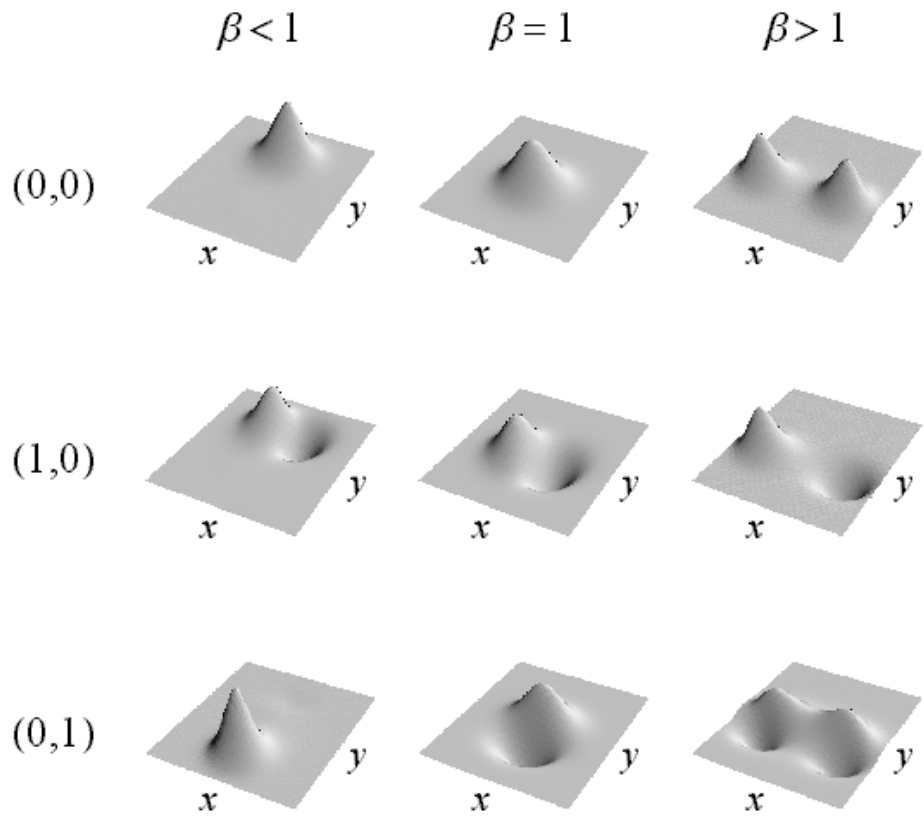

Figure 4. Hinds *et al*

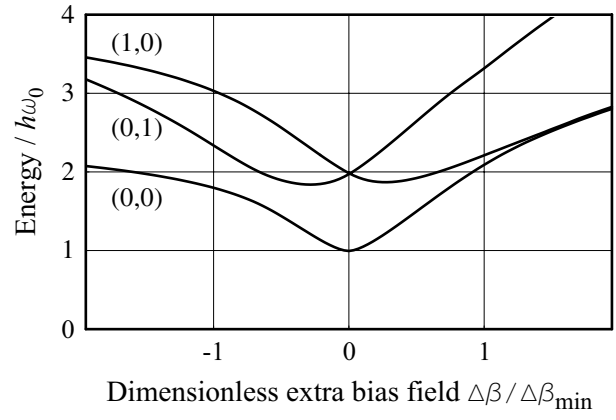

Figure 5. Hinds *et al.*